# Phase Noise and Frequency Accuracy in Crystal-less Wireless Edge Nodes

Xing Chen, *Member,* David D. Wentzloff, Senior *Member, IEEE*

*Abstract*—This paper presents a fundamental analysis connecting phase noise and long-term frequency accuracy of oscillators and explores the possibilities and limitations in crystal-less frequency calibration for wireless edge nodes from a noise-impact perspective. N-period-average jitter (NPAJ) is introduced as a link between the spectral characterization of phase noise and long-term frequency accuracy. It is found that flicker noise or other colored noise profiles coming from the reference in a frequency synthesizer is the dominant noise source affecting long-term frequency accuracy. An average processing unit embedded in an ADPLL is proposed based on the N-period-average jitter concept to enhance frequency accuracy in a 'Calibrate and Open-loop' scenario commonly used in low power radios. With this low-cost block, the frequency calibration accuracy can be directly associated with the reference noise performance. Thus, the feasibility of XO-less design with certain communication standards can be easily evaluated with the proposed theory.

## I. Introduction

THROUGHOUT the past decade, numerous efforts have been put into low power and low-cost wireless devices for ubiquitous inter-connected objects in the age of the Internet of Things, and there has been an increasing interest in the research of crystal-less (XO-less) radios for such applications. The majority of such designs are based on pulse modulations in wideband communications for the highest energy efficiency in the edge nodes [1-4]. Meanwhile, there is also plenty of research targeting to remove the XO in standard-compliant or compatible radio designs, such as in BLE [5-9], since an open-loop LC oscillator can offer enough phase noise (PN) performance [10,11] as long as the frequency is calibrated at the start of every packet transmission. In such systems where the RF oscillator's open-loop PN can easily satisfy the communication requirement, instead of locking all the time during RF transmission, the PLL's role can be shifted to a simpler 'calibration' scheme, relaxing the frequency requirements on the reference oscillator. In [5], the high-frequency XO is replaced with a 32kHz real-time clock (RTC) for frequency calibration and channel selection, while in [6], the XO is replaced with an FBAR resonator. [7] reported a BLE compatible XO-less transceiver by using a network-based frequency compensation from a Zigbee transmitter. [8,9] reported an XO-less BLE transmitter with clock recovery from GFSK-modulated BLE packets.

All this research shows the possibility of replacing the high-frequency XO in certain applications with other reference sources such as a received RF signal, low-frequency XO, and even a temperature-compensated RC oscillator. However, it still lacks a comprehensive analysis of how PN in the reference clock will impact the long term frequency accuracy (LTFA) in a 'Calibration and Open-loop' scheme incorporating all the reference cases mentioned above. Using jitter or instantaneous frequency errors to characterize the 'calibrated' frequency accuracy is inaccurate as the observation time span is different. Thus, we will introduce N-period Average jitter (NPAJ) to analyze this problem as a bridge between PN and LTFA.

The primary goal of this paper is to provide a fundamental analysis of the reference clock's noise impact on frequency accuracy. Section II summarizes the relationship among PN, NPAJ, and the widely used Allan Deviation (ADEV). Then the detailed impact of PN noise shaping on the LTFA will be analyzed. Section III will discuss how NPAJ is reflected in typical ADPLLs and how an embedded average processing unit (APU) would help increase the accuracy of frequency calibration. Section IV shows how different noise sources as frequency reference will affect the frequency accuracy and section V will draw the conclusion.

## II. Phase Noise and Long-Term Frequency Accuracy

### A. N-Period Average Jitter and Allan Deviation

PN has been evaluated and analyzed from various perspectives such as numerical methods [12-14], mathematical and physical understandings [15-18], circuit design considerations [19-25], and system-level requirements [10, 26], to name a few. However, its relation to the LTFA has rarely been discussed, and LTFA and PN analysis are normally treated as two separate subjects. But in the XO-less radio designs, this relationship becomes important as the noise from the reference increases and dominates the PN at low-frequency offsets. Its impact over time becomes more significant, and the analysis from PN's perspective is beneficial.

We start the analysis from the N-period jitter (NPJ), which is defined as the deviation between the phase jitter compared to its N-th previous value. The N-period jitter and PN's relation can be calculated from the autocorrelation of the NPJ and then taking the Fourier transform:

$$\sigma_{\tau_{pr}}^2(N) = \frac{2}{\pi^2 f_0^2} \int_0^\infty \mathcal{L}(f) \sin^2\left(\frac{\pi f N}{f_0}\right) df \quad (1)$$

Where $\sigma_{\tau_{pr}}(N)$ is the rms value of the NPJ, $\mathcal{L}(f)$ is phase noise and $f_0$ is the carrier frequency. From here we define the N-period average jitter (NPAJ) as the average value of the NPJ over the number of periods:

$$\tau_{prAVE}(N) = \frac{\tau_{pr}(N)}{N} \quad (2)$$

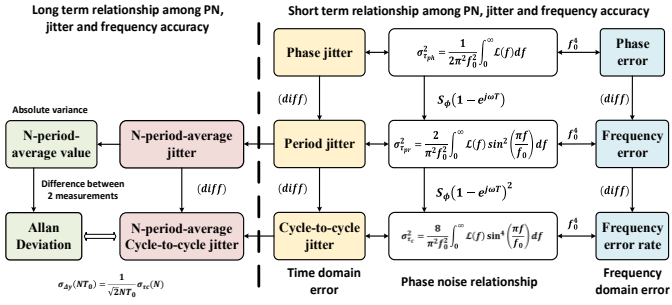

Fig. 1. Physical relations among phase noise, jitter, and frequency accuracy in both short term and long term

Which is still a jitter vector that can be related to frequency variations. It has a variance of:

$$\sigma^2_{\tau_{prAVE}} = \frac{2}{\pi^2 f_0^2 N^2} \int_0^\infty \mathcal{L}(f) \sin^2\left(\frac{\pi f N}{f_0}\right) df \quad (3)$$

Where $\sigma_{\tau_{prAVE}}$ is the rms value of the NPAJ.

On the other hand, ADEV [28-31] was originally calculated from the instantaneous frequency variation (IFV) over its nominal frequency and is defined as $y(t)$:

$$y(t) = \frac{\Delta f(t)}{f_0} = \frac{\tau_{pr}}{T_0} \quad (4)$$

Over the observation time, the average value of $y(t)$ is further calculated in evaluating the ADEV:

$$\overline{y_k} = \frac{1}{\tau} \int_{t_k}^{t_k+\tau} y(t) dt = \frac{\tau_{prAVE}(N)}{T_0} = \frac{\tau_{pr}(N)}{NT_0} \quad (5)$$

Where $\tau$ is the averaging time, and $t_k$ is the start of the sampling time. ADEV is defined as the variation of 2 consecutive samples of $\overline{y_k}$, the variance can be simplified as follows:

$$\sigma^2_{\Delta y}(NT) = \frac{1}{N^2 T_0^2} \frac{E\left[\left(t_p(k) - t_p(k+N)\right)^2\right]}{2} \quad (6)$$

Its relation to PN can be calculated using Wiener-Khintchin Theorem as:

$$\sigma^2_{\Delta y}(NT_0) = \frac{4}{\pi^2 N^2} \int_0^\infty \mathcal{L}(f) \sin^4(\pi f N T_0) df \quad (7)$$

Looking back at different jitter definitions, phase jitter is directly related to phase error over time, while period jitter is indirectly related to instantaneous frequency variations as it is the differential value between consecutive phase jitters. Similarly, cycle-to-cycle jitter is the difference between consecutive period jitters. Mathematically, cycle-to-cycle jitter is the derivative of period jitter and 2nd derivative of phase jitter. And ADEV is the N-period cycle-to-cycle jitter divided by N-period's time:

$$\sigma_{\Delta y}(NT_0) = \frac{1}{\sqrt{2}NT_0} \sigma_{tc}(N) \quad (8)$$

Where $\sigma_{tc}(N)$ is the rms value of the N-period cycle-to-cycle jitter. The relationship among jitter, PN, and LTFA has thus been clear to us and it is surprisingly simple. The widely used ADEV is simply the N-period cycle-to-cycle jitter with a coefficient related to the measuring time as shown in equation (8), and it is a derivative of NPAJ. Fig. 1 summarizes their relationships for short-term and long-term variations.

## B. Noise shaping's impact on long term frequency accuracy

The PN profile of a PLL can be viewed as a superposition of different noise sources shaped in various ways from a mathematical point of view. Moving from the lower frequency offset to the higher end, PN is firstly dominated by a flicker PN profile ($1/f^3$) from the reference clock, then to a nearly flat PN profile, and further to a white PN profile from the oscillator ($1/f^2$), and eventually to a flat noise floor. Different kinds of noise shaping have very different influences on NPAJ and ADEV over time. Assuming system bandwidth of $f_{BW}$, and for the flat PN profile, the NPAJ and ADEV from eq. (3) and eq. (7) can be simplified as:

$$\sigma^2_{\tau_{prAVE}} = \frac{\mathcal{L}_0}{\pi^2 f_0^2 N^2}\left[f_{BW} - \frac{\sin\left(\frac{2\pi f_{BW} N}{f_0}\right)}{\frac{2\pi N}{f_0}}\right] \quad (9)$$

$$\sigma^2_{\Delta y}(NT_0) = \frac{4\mathcal{L}_0}{\pi^2 N^2}\left\{\frac{3}{8}f_{BW} - \frac{\sin\left(\frac{2\pi N f_{BW}}{f_0}\right) - \frac{1}{8}\sin\left(\frac{4\pi N f_{BW}}{f_0}\right)}{\frac{4\pi N}{f_0}}\right\} \quad (10)$$

Where $\mathcal{L}_0$ denotes the flat PN level. It shows that with a flat PN profile, both the NPAJ and ADEV decrease with an increasing number of periods by $N$ as shown in Fig. 2.

For a white PN profile, generally assumed in free-running oscillators with white noise only, the PN follows:

$$\mathcal{L}(f) = \frac{\mathcal{L}_s f_s^2}{f^2} \quad (11)$$

Where $\mathcal{L}_s$ is the PN value sampled at $f_s$ offset to the carrier. The NPAJ and ADEV can be simplified as:

$$\sigma^2_{\tau_{prAVE}} \approx \frac{2\mathcal{L}_s f_s^2}{\pi f_0^3 N} Si(\infty) = \frac{\mathcal{L}_s f_s^2}{N f_0^3} \quad (12)$$

$$\sigma^2_{\Delta y}(NT_0) \approx \frac{2\mathcal{L}_s f_s^2}{\pi N f_0} Si(\infty) = \frac{\mathcal{L}_s f_s^2}{N f_0} \quad (13)$$

Where $Si(\infty) = \pi/2$ is the sine integral at infinity. The above two equations show that when only white noise is included, the NPAJ and ADEV will both decrease with time at the same rate following $\sqrt{N}$ as shown in Fig. 2.

And finally, the flicker PN profile can be written as:

$$\mathcal{L}(f) = \frac{\mathcal{L}_s f_s^3}{f^3} \quad (14)$$

Substituting (14) in (3), we have:

$$\sigma^2_{\tau_{prAVE}} = \frac{2\mathcal{L}_s f_s^3}{\pi^2 f_0^2 N^2} \int_0^\infty \frac{\sin^2\left(\frac{\pi f N}{f_0}\right)}{f^3} df \quad (15)$$

The above integral, however, will lead to non-converging results. Many theories have been developed in the past to explain the phenomenon of flicker noise, yet it is still obscure.

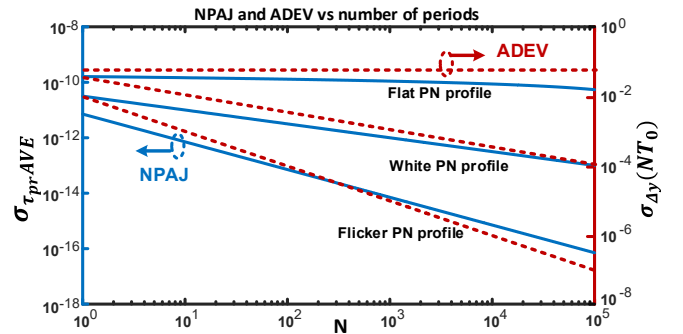

Fig. 2. NPAJ and ADEV vs number of averaging period with different PN profiles

The stochastic process that involves the flicker noise profile is actually non-stationary and simply applying the Wiener-Khintchin theorem is not 100% accurate. Thus, a frequency limit $f_{min}$ towards the carrier is necessary for all the approximated calculations. As explained in [17,27], with the finite observation time in one measurement compared to the lifetime of the object, the noise can be treated as an almost-stationary process as the non-stationary behavior is dominated by all its 'past'.

Thus, assuming $f_{min}$ as the lower integration boundary, (15) could be rearranged as:

$$\sigma^2_{\tau_{prAVE}} \approx \frac{\mathcal{L}_s f_s^3}{f_0^4}\left[3 - 2\gamma - 2\ln\left(\frac{2\pi N f_{min}}{f_0}\right)\right] \propto \ln\left(\frac{1}{N}\right) \quad (16)$$

Where $\gamma \approx 0.5772$ is the Euler-Mascheroni constant. In the same way, ADEV can be simplified as the following form:

$$\sigma^2_{\Delta y}(NT_0) \approx \frac{4\ln(2)\mathcal{L}_s f_s^3}{f_0^2} \quad (17)$$

Equation (16) shows that the NPAJ will almost keep unchanged yet still decrease with a rate of $\ln(1/N)$, while the ADEV is a constant in a flicker noise profile. It can be foreseen that if the PN has a spectrum like $1/f^\alpha$ and $\alpha > 3$, then both NPAJ and ADEV will increase with time.

### C. Limitation of averaging in LTFA

The above analysis shows that the time dependence of both NPAJ and ADEV to characterize an oscillator's LTFA varies with different PN profiles. As discussed above, frequency variations due to flat and white PN profiles can be minimized with the increase of averaging time while flicker noise will stop that trend from a statistical point of view. With different measurement (averaging) time, the contribution of different PN profiles to frequency variation varies significantly. Intuitively, assuming the PN sampling point in (12) and (17) is the same as the flicker noise corner for a free-running oscillator, the NPAJ due to white PN and flicker PN can be compared as follows:

$$\frac{\sigma^2_{\tau_{prAVE_{flicker}}}(N)}{\sigma^2_{\tau_{prAVE_{white}}}(N)} = \frac{f_c}{f_0} N \left[3 - 2\gamma - \ln\left(\frac{2\pi f_{min} N}{f_0}\right)\right] \quad (18)$$

This shows that the jitter contribution from flicker noise is only affected by the flicker noise corner and averaging time N. Fig. 3 shows the ratio of flicker PN jitter to white PN jitter in the NPAJ over time with different flicker noise corners plotted in log scale. It can be seen that when N=1, white PN contributes the majority of period jitter variations while with the increase of averaging time, flicker PN becomes the dominant noise source.

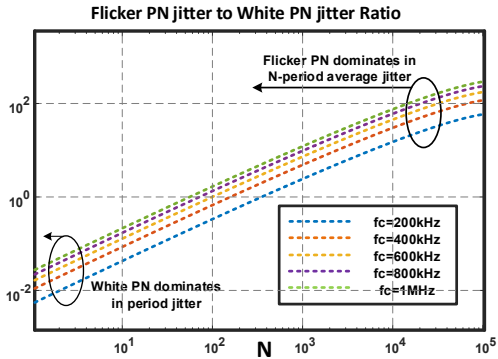

Fig. 3. Ratio of flicker PN jitter and white PN jitter with $f_0 = 1GHz$ and different flicker noise corners

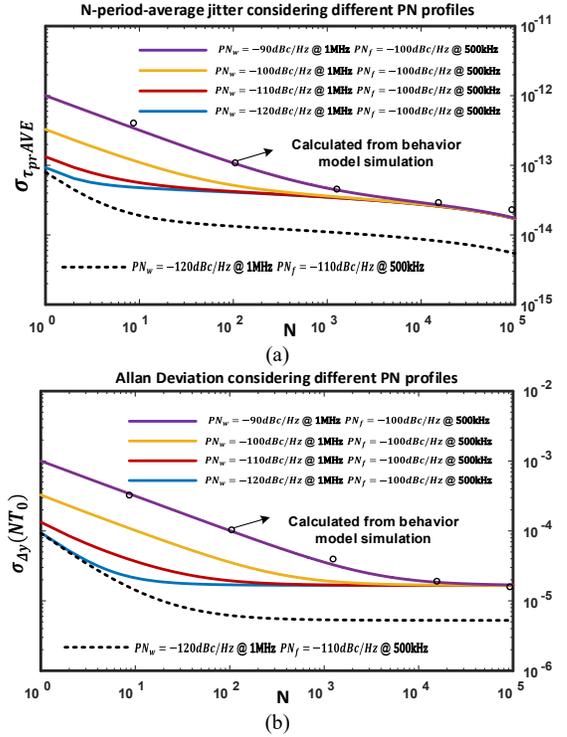

Fig. 4. NPAJ (a) and ADEV (b) of typical open loop oscillator including flat, white and flicker PN profiles in different levels.

So eventually, the frequency synthesizer's LTFA is determined by the reference clock's noise performance and limited by its flicker noise component no matter how much more noise was added by other noise sources in the PLL. The reference clock's PN is lowpass filtered and upconverted by the frequency multiplication ratio, thus, in the long term, the frequency stability of the PLL output is a direct reflection of the reference clock's frequency stability. Fig. 4 shows the simulated NPAJ and ADEV over time with different PN profiles. In the low-frequency offset, their PNs are all dominated by the same flicker PN. And after certain periods of time, all the PN profiles converge at the same level.

The above analysis and simulations show 3 important features in frequency averaging:
1. Flicker noise has a small impact compared to white noise in the short-term jitter while it has a significant impact in the long-term frequency accuracy, according to the flicker noise corner (eq. 21).
2. It is the flicker noise from the reference clock that dominates the long-term frequency accuracy performance, and all other noise components in higher frequency offsets can be averaged out.
3. In XO-less applications, averaging could help to improve the accuracy of frequency calibration to the physical limit set by the reference clock's flicker noise corner.

### III. EMBEDDED AVERAGE PROCESSING UNIT IN ADPLL FOR FREQUENCY CALIBRATION

As discussed above, all noise components other than the flicker noise from the reference clock could be eventually canceled out through averaging in frequency calibration. Thus, in order to do frequency calibration with a noisy reference clock, an embedded average processing unit (APU) in an ADPLL is proposed [8, 9], which has been demonstrated to be able to do accurate frequency calibration using a GFSK modulated signal. It is based on the traditional type I divider-less ADPLL and the APU is embedded in the digital loop filter and performs a windowed averaging algorithm of the digital control word (DCW).

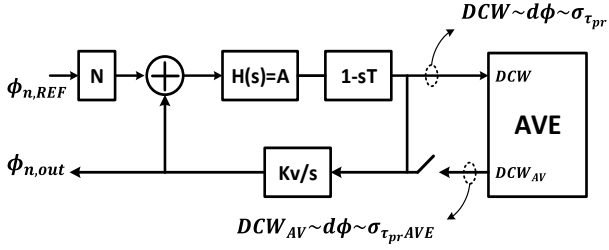

Fig. 5. Continuous time behavior model of the proposed frequency calibration circuit considering only reference noise source

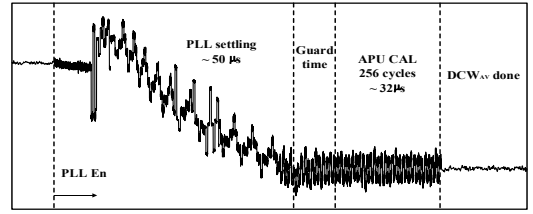

Fig. 6. Measured transient frequency response of PLL settling and frequency calibration from the APU using recovered RF signal as reference

The reason that averaging the DCW would be useful is that it is a direct reflection of the phase error difference between the reference phase and the output phase with one loop delay in the PLL. Thus, the DCW corresponds to the period jitter of the whole system while the $DCW_{AV}$ correspond to the NPAJ defined in previous sections. Fig. 5 shows the corresponding continuous-time behavior model of the ADPLL with embedded APU proposed in [8]. In the typical 'calibrate and open-loop' scheme, using $DCW_{AV}$ rather than DCW will result in a much more accurate 'releasing frequency' when open-loop [8]. During the locking status of a PLL, the DCW in real-time is a random process following a Gaussian distribution if no significant spurs exist. The 'releasing frequency' could be anywhere within $\pm 3\sigma_{\tau p}$ of the carrier frequency. While the 'releasing frequency' by using $DCW_{AV}$ is within $\pm 3\sigma_{\tau p_{AVE}}$ of the carrier. Fig. 6 shows the measured transient response of the ADPLL with APU reported in [8] illustrating how intuitively the APU helps to improve frequency calibration accuracy using the recovered RF signal. With enough time in the average process, $\sigma_{\tau p_{AVE}}$ could be much smaller than $\sigma_{\tau p}$ depending on the PN profile of the reference itself. As analyzed in section II, longer averaging time would always help with the NPAJ even when it exceeds the time constant associated with the flicker noise corner, which corresponds to $N_c = \ln(2) f_0/4f_c$, where $f_0$ is the carrier frequency and $f_c$ is the flicker noise corner. However, considering practical issues such as memory size, power, and area of the APU, a total averaging time close to $N_c$ is more efficient. In the frequency domain, this embedded APU offers a simple digital low pass filter to DCW [8], and since it's clocked by the reference clock, the power consumption and overall cost of this block are much lower compared to other PN filtering techniques in recent publications [33].

## IV. DIFFERENT TYPES OF REFERENCES AND XO-LESS FEASIBILITY

### A. Different types of reference noise

As mentioned in section I, several reference clocks that have been evaluated in recent publications could replace the high-frequency XO in wireless edge nodes: RTC, RC oscillator, and recovered RF signal. RTC is widely used in SoCs for clocking, but strictly speaking, it mostly uses a crystal as well and it is an accurate reference source operating at a low frequency. RC oscillators, on the other hand, are relatively noisy reference sources. And to use an RC oscillator as a reference for frequency calibration, its frequency and temperature dependence shall be pre-characterized with an accurate temperature sensor included in the design as well. The RF clock recovery case has a rather complicated PN profile since the RF signal could be modulated. In BLE for example, the RF signal seen by the receiver on the wireless edge node is GFSK modulated. The frequency variation depends on the data packet, but the center frequency can be considered accurate as it is regulated by the XO in the transmitter. From the noise perspective, the RTC represents a low PN profile, the RC oscillator represents a high PN profile, and the recovered RF signal represents a complex PN profile depending on its modulation format. Overall, the modulation would only affect the PN in a relatively high-frequency offset. And in all cases, it is still the flicker noise after up-conversion with its multiplication ratio $N = f_{RF}/f_{ref}$ that matters. Fig. 6 shows PN profiles of the 3 cases where the RTC is referred to an XO design from [34] and the corresponding behavior simulation results. The RC oscillator is referred to [35] and the RF signal shows both single tone and GFSK modulated signals. The corresponding simulated frequency accuracy with their approximated averaging number of periods are also shown.

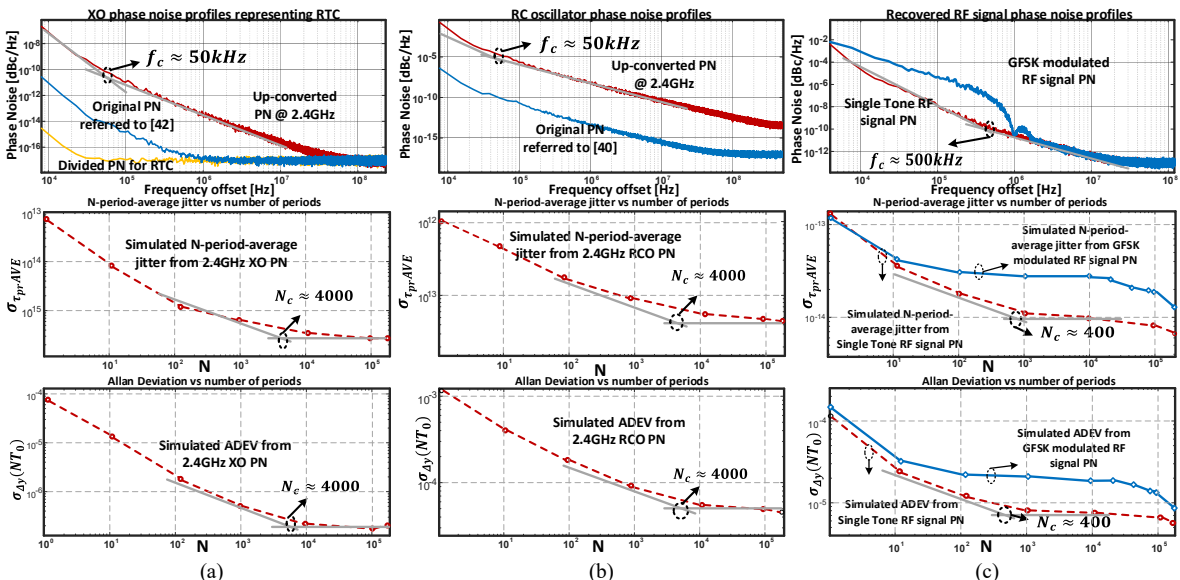

Fig. 7. Simulated phase noise impact on NPAJ and ADEV for different noise sources: (a) RTC, (b) RC oscillator, (c) Recovered RF signal

Fig. 7 (a) shows a typical RTC design divided from a high frequency XO oscillator. For previous 32.768 kHz RTC designs, it's rare to report phase noise performance at such a low frequency. So, it is chosen as an example RTC for its 'excellent noise performance'. The blue line is its original measured phase noise at 38.4MHz and the yellow line is the divided RTC noise performance. To be fair for all 3 cases, the red lines in all PN plots are upconverted to the same center frequency at 2.4GHz. The corresponding NPAJ and ADEV of that RTC reference are shown below in Fig. 7 (a) as well. As the flicker noise corner of the RTC is around 50 kHz, the resulting number of periods where ADEV turns flat is close to 4000. The behavior simulation using such a PN profile shows a similar result. Fig. 7 (b) represents an excellent RC oscillator design. [34] reported its phase noise performance at 10MHz, which is redrawn in the blue line. The red is the up-converted PN at 2.4GHz as well. As can be seen from the NPAJ and ADEV plots, they also have a $N_c \approx 4000$ as the flicker noise corner is around 50 kHz as well. But since its noise performance is much worse than the RTC case in Fig. 7 (a), both jitter performance and ADEV are around 30 times worse than the RTC case. The recovered RF signals were shown in Fig. 7 (c). The single tone RF signal is directly coming from an open-loop LC oscillator design at 2.4GHz while the other is GFSK modulated representing a general BLE packet. It has a 1Mbps data rate and 500 kHz frequency deviation. As seen from the NPAJ and ADEV plots, the two cases will converge with the increased number of periods, although the GFSK modulated case deviates from the single tone case in the middle. For the single tone case, the flicker noise corner is around 500 kHz and both the jitter and ADEV turn flat at around 400 periods. The jitter and ADEV plot for the GFSK modulated case could be more complicated, since with the limited behavior simulation results, those regions have been masked by the blue line.

### B. A Bluetooth Low-Energy example

The above simulation results further show that it is the flicker noise that eventually defines the frequency accuracy after calibration. Fig. 8 shows the relation between flicker PN and corresponding ADEV at 2.4 GHz. It can be seen that the RC oscillator from [35] shows a $2^{-5}$ ADEV while the XO and RTC would offer around $6^{-9}$ ADEV. The RF signal, no matter modulated or not, would be able to offer better than $2^{-6}$ ADEV with enough averaging time. If we take BLE as an example, as it has been proven that an open-loop LC oscillator can meet the BLE phase noise requirement, we could specify the phase noise requirement of its reference from the analysis above. According to the Bluetooth standard [36], it requires $\pm 150\ kHz$ frequency offset, which corresponds to a 60ppm frequency accuracy requirement in Allan Deviation. It can be seen that any oscillator with minimum ADEV<60ppm (flat region of the ADEV plot) with an accurately characterized temperature coefficient can be used as a frequency reference in XO-less BLE edge nodes.

## V. CONCLUSION

In this paper, phase noise and frequency accuracy has been analyzed for XO-less wireless edge nodes. The relationship between PN and LTFA such as NPAJ and ADEV has been analyzed considering different PN profiles. It has been found that flicker noise plays an important role in defining how accurate the frequency of an oscillator could be over a long period of time while all other noise sources could be filtered simply by averaging. The APU embedded in ADPLL proposed in recent XO-less BLE designs is also evaluated based on the NPAJ analysis. Although previously demonstrated, the APU is further simulated using different reference clocks through a behavioral model. The results show that although RTC offers significantly better performance compared to recovered RF signal and RC oscillators, the latter two can actually be used in frequency calibration in BLE. Theoretically, a state-of-the-art RC oscillator could be used as a crystal replacement in the low-cost wireless edge nodes. However, this paper only considers the PN's impact on frequency accuracy without taking PVT variations into account. The recovered RF signal, which has been demonstrated in recent publications, on the other hand, is more promising in the XO-less edge node designs as the incoming signal from a base station is already characterized. The modulation wouldn't necessarily affect the frequency calibration result providing enough time for the average process.

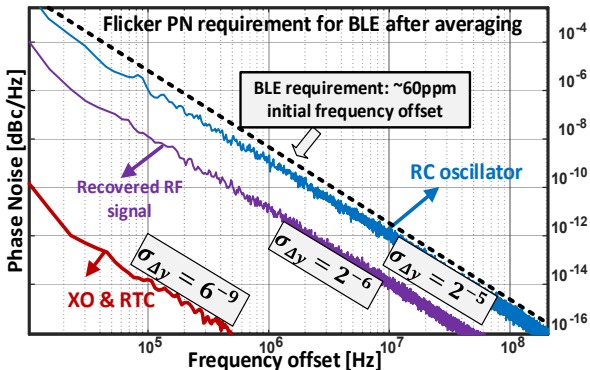

Fig. 8. Typical range of Allan Deviation according to flicker noise of different noise sources, and requirement for the reference clock in BLE applications